# A Hybrid (Al)GaAs-LiNbO$_3$ Surface Acoustic Wave Resonator for Cavity Quantum Dot Optomechanics

Emeline D. S. Nysten[1], Armando Rastelli[2], Hubert J. Krenner[1*]

[1] Lehrstuhl für Experimentalphysik 1 and Augsburg Centre for Innovative Technologies (ACIT), Universität Augsburg, Universitätsstraße 1, 86159 Augsburg, Germany

[2] Institute of Semiconductor and Solid State Physics, Johannes Kepler Universität Linz, Linz Institute of Technology, Altenbergerstraße 69, 4040 Linz, Austria

[*] corresponding author: hubert.krenner@physik.uni-augsburg.de

**Abstract**

A hybrid device comprising a (Al)GaAs quantum dot heterostructure and a LiNbO$_3$ surface acoustic wave resonator is fabricated by heterointegration. High acoustic quality factors $Q > 4000$ are demonstrated for an operation frequency $f \approx 300 \text{ MHz}$. The measured large quality factor-frequency products $Q \times f > 10^{12}$ ensures the suppression of decoherence due to thermal noise for temperatures exceeding $T > 50 \text{ K}$. Frequency and position dependent optomechanical coupling of single quantum dots and the resonator modes is observed.

**Main Text**

Elastic waves and acoustic phonons are known to couple to literally any excitation in condensed matter. This unique property makes them ideally suited for the design and realization of hybrid quantum systems[1]. Recently, surface acoustic waves (SAWs)[2], i.e. surface-confined elastic waves shifted back into focus of this active field of research. These coherent radio frequency (rf) phonons enable versatile quantum transduction[3] and dynamic, non-adiabatic control of quantum systems[4]. In experiment, the SAWs have been employed for the coherent control of superconducting qubits in the single phonon limit[5], on-chip quantum state transfer between superconducting qubits by single SAW quanta[6], single electron spin transfer between electrostatic quantum dots (QDs)[7], coherent acoustic control of single spins[8,9], defect centers[10,11] and optically active QDs[12–15]. Optically active, epitaxial QDs exhibit distinct advantages for the design of hybrid quantum architectures. Their emission wavelength can be tuned by chemical composition and size[16] or post-growth by external parameters such as electric or magnetic fields[17] or strain[18]. In addition, the tunable coupling can be achieved between excitons in multi-dot architectures[19,20] and excitons and optical modes in photonic systems[21,22]. In SAW technology and nonlinear optics, Lithium Niobate (LiNbO$_3$)[23] is the substrate material of choice because of its high electromechanical coupling coefficient $K^2 \approx 5\,\%$ ($K^2 \approx 0.07\,\%$ for GaAs) and $\chi^{(2)}$ ($\chi^{(2)} = 0$ for GaAs) optical nonlinearity, respectively. Because, LiNbO$_3$ does not provide any type of high-quality qubit system, the design and fabrication of hybrid quantum devices requires its heterointegration with other materials. Here, we report on the realization of a hybrid SAW resonator device comprising a SAW cavity defined on a LiNbO$_3$ substrate and epitaxially grown optically active QDs. We demonstrate optomechanical coupling of single QDs to the phononic modes of the resonator. This coupling is determined by the local amplitude of the acoustic field at the QD's position. Interestingly, the QD's optomechanical response exhibits a richer spectrum than the electrically determined resonator properties, opening new directions for future explorations employing our hybrid device.

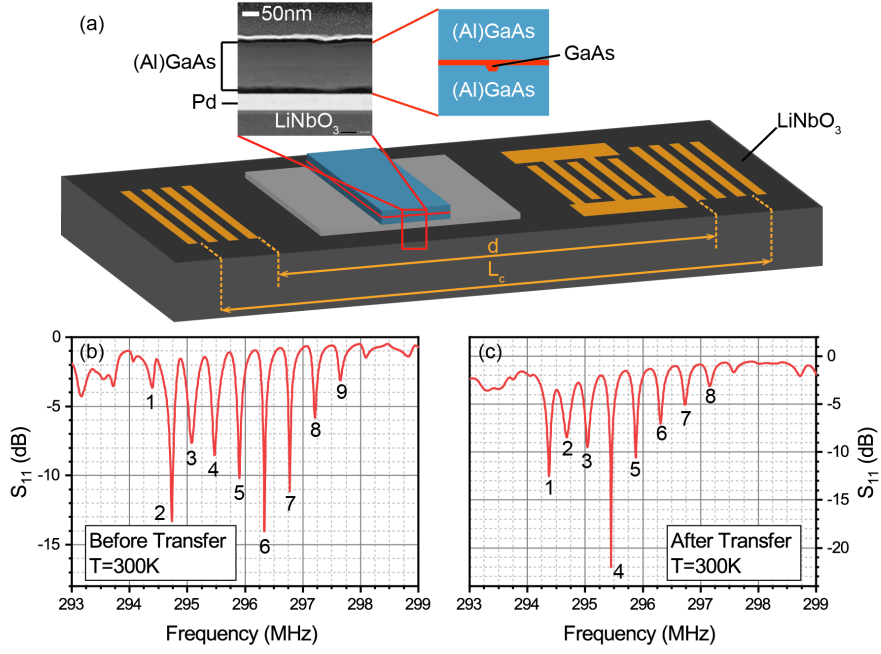

Figure 1 – (a) Schematic of the hybrid device comprising a LiNbO$_3$ single port SAW resonator and a (Al)GaAs heterostructure containing a single layer of QDs on a Pd adhesion layer. The inset shows a TEM image of the LiNbO$_3$-Pd-(Al)GaAs stack. Measured room temperature rf reflectivity ($S_{11}$) of the SAW resonator before (b) and after (c) heterointegration of the QD layer.

Our device is fabricated by heterointegration of a (Al)GaAs heterostructure containing a single layer of droplet etched QDs onto a conventional single port LiNbO$_3$ SAW resonator device[24]. A schematic of our device is shown in Figure 1 (a). The SAW resonator is patterned onto an oxygen-reduced 128° rotated Y-cut LiNbO$_{3-x}$ substrate. The resonator is formed by two metallic floating electrode acoustic Bragg-reflectors (150 fingers, aperture $a = 350$ μm, nominal mirror separation $d = 4522$ μm) and is aligned along the X-direction. The phase velocity of the SAW is $c_{SAW,0} = 3990$ m/s along this direction. The nominal acoustic design wavelength and frequency are $\bar{\lambda}_n = 13.3$ μm and $\bar{f}_n = 300$ MHz, respectively. The resonator is excited by applying an electrical rf signal of frequency $f_{rf}$ to a 20 finger pairs interdigital transducer (IDT). The acoustic Bragg mirrors and the IDT are patterned during the same electron beam lithography step and finalized using a Ti (5 nm) / Al (50 nm) metallization in a lift-off process. The IDT is positioned off-center, close to one Bragg mirror and the large open area is used for the heterointegration of the III-V compound semiconductor film. Figure 1 (b) shows the rf reflectivity of our resonator device measured with the IDT at $T$ = 300 K. In this spectrum we can identify nine pronounced phononic modes, which are consecutively numbered. The measured complex reflection $S_{11}(f)$ can be fitted by[25]

$$\tilde{S}_{11}(f) = \frac{(Q_e - Q_{i,n})/Q_e + 2iQ_{i,n}(f - f_n)/f}{(Q_e + Q_{i,n})/Q_e + 2iQ_{i,n}(f - f_n)/f}.$$

*Equation 1*

In this expression $Q_{i,n}$ and $Q_e$ denote the internal and external quality factor of mode $n$ and external circuit, respectively. $f_n$ is the resonance frequency of the $n$-th mode. We find a mean $\bar{Q}_i = 2900 \pm 700$ ($\overline{\Delta f} = 100 \pm 20$ kHz) and $\bar{f}_n = 296$ MHz at room temperature. The given values are the mean of the distribution and their standard deviation of the mean. The full analysis is included in the supplementary material. These modes are split by the free spectral length $FSR_{empty} = 416 \pm 25$ kHz. This value corresponds to a cavity roundtrip time of $T_c = \frac{1}{FSR} = 2.41 \pm 0.15$ µs and a resonator length $L_c = \frac{c_{SAW,0}}{2FSR} = 4800 \pm 50$ µm. The penetration length of the acoustic field into the mirror is given by $L_p = w/|r_s| = 145$ µm, where $w = 3.3$ µm is the width of the fingers of the mirror and $r_s = 0.023$ is the reflectivity coefficient of one finger[25,26]. Using the lithographically defined $d$, we calculate a resonator length $d + 2L_p = 4810$ µm, which agrees well with the value derived from the experimental data. The heterointegration is realized by epitaxial lift-off and transfer onto a 50 nm thick and 3000 µm long Pd adhesion layer[27–31]. The heterostructure was grown by molecular beam epitaxy and consists of a 150 nm thick $Al_{0.33}Ga_{0.67}As$ membrane with a layer of strain-free GaAs QDs [32] in its center. The membrane was heterointegrated onto the $LiNbO_3$ SAW-resonator by epitaxial lift-off and transfer. In essence, the QD heterostructure is released from the growth substrate by selective etching of an $Al_{0.75}Ga_{0.25}As$ sacrificial layer using hydrofluoric acid. In the next step, the membrane is transferred onto the Pd adhesion layer in the center of the SAW resonator and a rectangular piece is isolated by wet-chemical etching. Further details can be found in Ref. [29] and the supplementary material. A transmission electron microscope (TEM) image of the $LiNbO_3$-Pd-(Al)GaAs stack is shown in Figure 1 (a). The semiconductor membrane is laterally placed in the center of the resonator. After transfer, the membrane is etched to obtain straight edges and, thus, reduce scattering losses. The final membrane is 215 µm wide and extends over the full width of the resonator. Further details on the heterostructure and an optical microscopy image are included in the supplementary material. The resonator mode spectrum after transfer recoded at $T$ = 300 K is shown in Figure 1 (c) and is analyzed using Equation 1. The full analysis is also part of the supplementary material. By comparing these data to those before transfer we find that the mode spectrum and $FSR$ remain approximately constant within the experimental error at $\bar{f}_n = 295.8$ MHz and $FSR_{hybrid} = 406 \pm 22$ kHz. The corresponding cavity roundtrip time is $T_{c,hybrid} = 2.46 \pm 0.13$ µs. Most importantly, high internal quality factors of $\bar{Q}_i = 2500 \pm 300$ ($\overline{\Delta f} = 120 \pm 15$ kHz) are preserved after transfer, which is of highest relevance for strong phonon-exciton coupling. Furthermore, all experimental data is well reproduced by finite element modelling (FEM) detailed in the supplementary material. For example, the experimental change of $T_c$ after heterointegration of $\Delta T_c = 50$ ns is in excellent agreement with $60$ ns predicted by FEM. Furthermore, the reduction of the effective phase velocity in the hybridized region to $c_{SAW,eff} = 3889$ m/s gives rise to a spectral shift of the mode spectrum of $\Delta f_n = 10.5$ MHz to lower frequencies. Note, that according to these calculations the absolute mode index changes from $n_{abs,0} = n + 707$ of the bare resonator to $n_{abs,eff} = n + 726$, for the hybrid device.

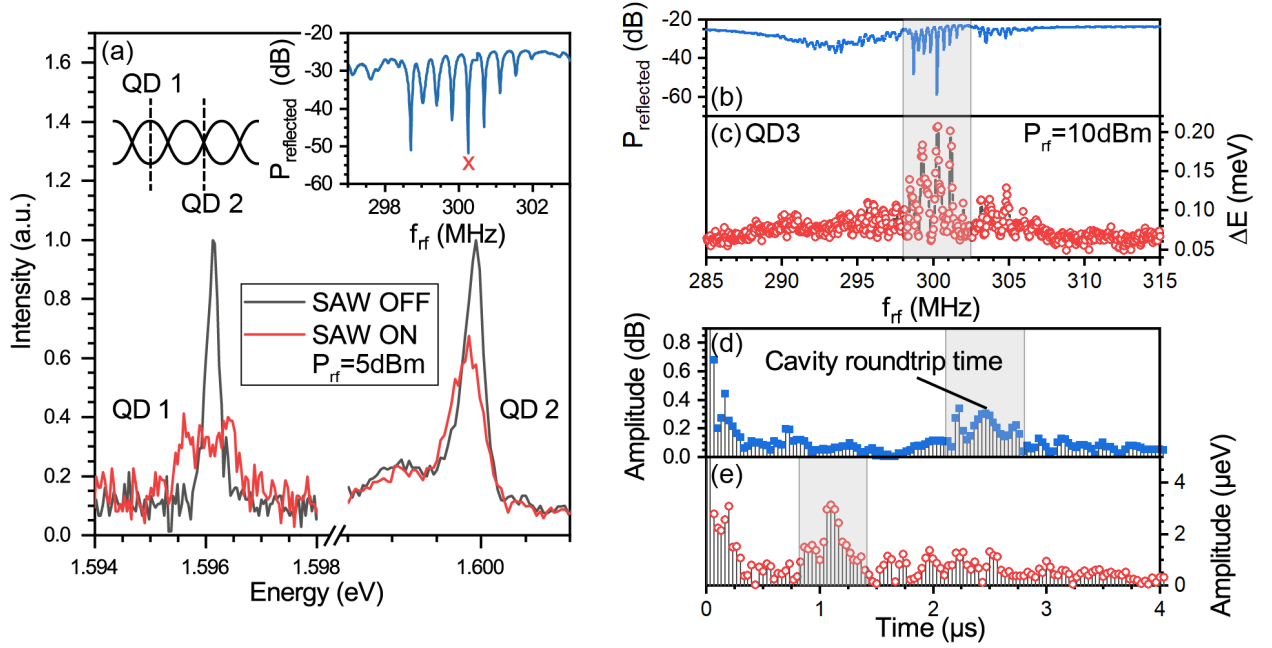

*Figure 2 – (a) Low temperature emission spectra of two QDs inside the SAW resonator without (black lines) and with (red lines) $f_{rf} = 300.25\ MHz$ applied with $P_{rf} = 5\ dBm$ to the IDT. This applied frequency is resonant to the $n = 5$ mode marked in the measured reflected power spectrum of the resonator (inset). QD1 (QD2) is located close or at an antinode (a node) of the cavity field as shown by the schematic. (b) Reflected power spectrum and (c) simultaneous optomechanical response of QD3. (d, e) FFT of the data in (b) and (c) showing a clear signature at $T_c$ and $T_c/2$, respectively.*

Next, we investigate the optomechanical coupling of single QDs to the phononic modes of the resonator in Figure 2. The dominant coupling mechanism is deformation potential coupling, i.e. the modulation of the semiconductor's bandgap by hydrostatic pressure, i.e. normal stress[15]. The Pd adhesion layer effectively shortens any electric field[29,30]. Thus, Stark effect modulation, which becomes dominant at high SAW amplitudes[33], is to be strongly suppressed. We measure the optomechanical response at low temperatures ($T = 10$ K) by time and phase averaged micro-photoluminescence spectroscopy[15]. Importantly, we record the reflected electrical power ($P_{reflected}$) at every step, i.e. for any combination of electrical frequency ($f_{rf}$) and power ($P_{rf}$) applied to the IDT. Thus, we eliminate potential sources of errors for instance due to temperature related drifts of the mode spectrum. Further details can be found in the supplementary material. In essence, the detected lineshape is a time-average of the dynamic optomechanical modulation of the unperturbed, Lorentzian QD emission line[33]. In a first step, we apply $P_{rf} = 5$ dBm to the IDT at $f_5 = 300.25$ MHz. The measured $P_{reflected}$ is plotted as a function of $f_{rf}$ in the inset of Figure 2 (a). The main panel shows emission spectra of two QDs, QD1 and QD2 with (red) and without (black) the SAW resonating in the cavity. The two QDs are separated by $\simeq 21\ \mu m \simeq 1.6\ \lambda_{SAW}$ along the cavity axis and exhibit completely dissimilar behavior. While QD1 shows a pronounced broadening when the SAW is generated, the lineshape of QD2, apart from a weak reduction of the overall intensity remains unaffected. These types of behaviors are expected for QDs positioned at an antinode (QD1) or node (QD2) of the acoustic cavity field. From the observed optomechanical responses, we infer that QD1 and QD2 are located at an antinode or node of the mode, respectively, as illustrated by the schematic. In a second step, we keep the optical excitation fixed and scan the radio frequency $f_{rf}$ applied at a constant power level over wide range of frequencies $285 - 315$ MHz and record emission spectra of a single QD (QD3). These data are fitted with a time-integrated, sinusoidally modulated Lorentzian[29,34] of width $w$ and amplitude $A$.

$$I(E) = I_0 + f_{rf} \frac{2A}{\pi} \int_0^{1/f_{rf}} \frac{w}{4 \cdot \left(E - \left(E_0 + \Delta E \cdot sin(2\pi \cdot f_{rf} \cdot t)\right)\right)^2 + w^2} dt.$$

*Equation 2*

In Equation 2, $E_0$ and $\Delta E$ denote the center energy of the emission peak and the optomechanical modulation amplitude due to the time-dependent deformation potential coupling. From our established FEM modelling[29] we obtain an optomechanical coupling parameter[14,15] $\gamma_{om} = 2500$ µeV/nm. Moreover, the measured $\Delta E$ per repetition of the IDT pattern is enhance by at least a factor of 2 when compared to the previously studied delay line device[29]. Figure 2 (b) and (c) show the simultaneously recorded reflected rf power ($P_{reflected}$) and $\Delta E$ as a function of $f_{rf}$. Clearly, QD3 exhibits a series of strong optomechanical modulation peaks at frequencies at which pronounced cavity modes are observed (grey shaded area). This observation of a pronounced coupling to resonator modes is a first direct evidence of cavity enhanced coupling between SAW phonons and the exciton transition of a single QD. However, the detected optomechanical response, $\Delta E(f_{rf})$, of QD3 exhibits noticeably less peaks than $P_{reflected}$. We Fast Fourier transform (FFT) $P_{reflected}(f_{rf})$ and $\Delta E(f_{rf})$ to obtain time domain information. The result of these Fourier transform is plotted in Figure 2 (d) and (e). In the FFT of $P_{reflected}(f_{rf})$ in (d), clear peak at $t = 2.4 \pm 0.05$ µs can be identified, which matches exactly the cavity roundtrip time, $T_c = 2.41$ µs of the SAW resonator derived from the measured $FSR$. In contrast, the FFT of $\Delta E(f_{rf})$ in (e) shows a clear signal at $t = 1.1 \pm 0.1$ µs $\simeq T_c/2$. This apparent halving of the roundtrip time, i.e. doubling of the $FSR$, in the dot's optomechanical response provides first direct evidence that coupling occurs only to every second cavity mode.

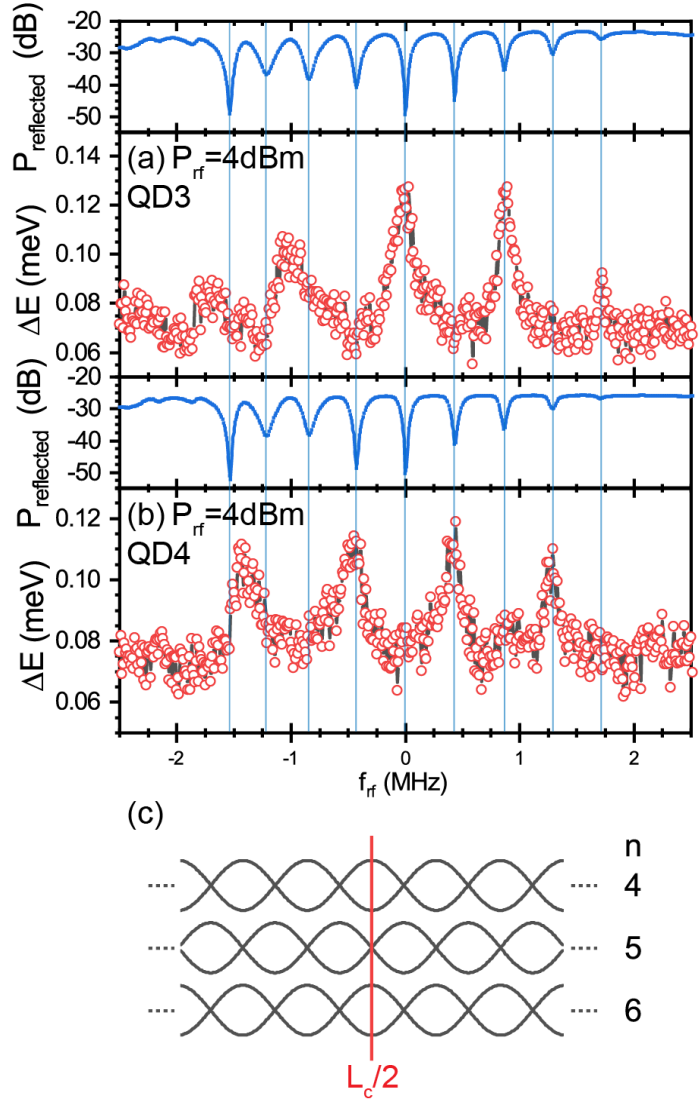

*Figure 3 – rf-dependent optomechanical response of QD3 (a) and QD4 (b) measured at $T = 10\,K$. Upper panels: Reflected rf power $P_{reflected}$. Main panels: Optomechanical response $\Delta E\,(f_{rf})$. (c) Schematic of the acoustic field in the center of the resonator for the modes detected in the experimental data above.*

We continue studying this mode index selective coupling in more detail. In Figure 3 we investigate the $f_{rf}$-dependence of the optomechanical response of QD3 and another different dot, QD4, in (a) and (b), respectively. The main panels show the optomechanical modulation amplitude $\Delta E$ derived from best fits of Equation 2 and the upper panels the simultaneously measured $P_{reflected}$. All data are plotted as a function of the frequency shift with respect to the center mode $n = 5$. From these electrical data we obtain the low temperature value of the mean quality factor $\bar{Q} = 4430 \pm 1560$ an increase by a factor of $\approx 1.75$ compared to the room temperature value. Furthermore, the resulting high quality factor-frequency product ($Q \times f$), a commonly used figure of merit used to compare mechanical resonators[35]. Here we obtain $\bar{Q} \times f = (1.33 \pm 0.48) \cdot 10^{12}$, which in the quantum realm has to be compared to the thermal energy. In our case, the SAW phonon energy of $\approx 1.25\,\mu eV$ is less than $k_B T \approx 250\,\mu eV$ and the maximum number of coherent oscillations, i.e. number of coherent operations possible, is limited by thermal decoherence to $Q \times f \times \frac{h}{k_B T}$. At $T = 10\,K$, thermal decoherence limits the number of coherent oscillations[36] in the unloaded hybrid resonator device to $\bar{Q} \times f \times \frac{h}{k_B T} \approx 6$. A single oscillation will be preserved even up to 50 K. QD3 shows a strong optomechanical response when modes with odd mode index $n = 5, 7$ are excited. In contrast, QD4 couples to modes with even index $n = 4, 6, 8$. The width of these resonances corresponds to an optomechanically detected quality factor $\overline{Q_{QD}} = 1730 \pm 420$ ($\overline{\Delta f_{QD}} = 890 \pm 30 \text{kHz}$). This decrease compared to the electrically measured value may arise from imperfections during heterointegration, e.g. inhomogeneous bonding and roughness and misalignment of the etched edges. Moreover, the splitting between modes which optomechanically couple to the QD is doubled compared to the $FSR$ measured electrically, and consequently, the corresponding time is half of the cavity roundtrip time. The alternating coupling behavior can be understood well considering the position of nodes and antinodes of the acoustic fields of different modes in the center of resonator. The qualitative profiles of the $n = 4, 5$ and 6 are shown in Figure 3 (c). Clearly, modes with even (odd) index exhibit nodes (antinodes). Thus, a single QD positioned at nodes or antinodes can be selectively coupled to modes with either even or odd mode index, and QD3 and QD4 are two representative examples for each case. This simple picture applies well to modes $n \geq 4$. For $n \leq 3$, a more complex behavior is observed. For QD4, we observe a strong optomechanical response at $(f_2 + f_1)/2$ and for QD3 similarly at $(f_3 + f_2)/2$. At the same time no optomechanical response is detected for $n = 1$ and $n = 3$, and for $n = 2$, expected for QD3 and QD4, respectively. For linear optomechanical coupling between the resonating SAW and the QD, a pronounced optomechanical response should be only detected only when the driving rf signal is in resonance with modes of odd (QD3) and even (QD4) index. Moreover, no optomechanical response should be detected when $f_{rf}$ is detuned from a cavity mode because the large cavity efficiently rejects phonons in this case. Despite the fact that the acoustic power injected in the resonator remains finite in a real, lossy device like ours, the observed behavior cannot be attributed to simple, linear coupling mediated by the deformation potential. In the supplementary material, we show data from another QD5. This dot is heterointegrated in a different, nominally identical SAW resonator. QD5 shows identical optomechanical response as QD3, thus these data corroborate that the observations made are reproducible for this type of device. This in particular excludes that the observation can be solely attributed to local imperfections. As shown in Figure 4 (a), the optical linewidth of the QD[37] $\Gamma_{QD} \sim 1.25$ GHz, fully covers the entire phononic mode spectrum. Thus, coupling between all phononic modes can occur. Figure 4 (b-d) compares the measured optomechanical response of QD3 for three different $P_{rf}$. The full analysis is included in the supplementary information. The reflected rf power is given as a reference in the upper panels. As $P_{rf}$ increases the optomechanical modulation amplitude $\Delta E$ of QD3 increases and, moreover, new features develop, which are not observed for low $P_{rf}$ in Figure 3 (b). The reflected electrical power spectra in the upper panels do not show similar pronounced changes which indicates that there is no strong backaction and transduction to the acoustic domain. Most notably, at the highest power level applied to the IDT, i.e. maximum number of phonons injected into the resonator, $P_{rf} = 16$ dBm, we observe clearly resolved new features at $(f_4 + f_3)/2$ and $(f_6 + f_5)/2$. Again, in an ideal device and for coupling mediated by the deformation potential, these features are completely unexpected. We study the dependence of the optomechanical modulation on $P_{rf}$. Such $f_{rf}$-scans were recorded for nine different $P_{rf}$. We extracted the maximum of the optomechanical modulation amplitude $\Delta E_{max}$ at $(f_3 + f_2)/2$ (1, black), $f_5$ (2, red) and $f_7$ (3, blue). The data is plotted as symbols in logarithmic representation as a function of $P_{rf}$ in Figure 4 (e) to identify power law dependencies. The lines in Figure 4 (e) are linear fits to the data from which we are able to determine the power law for the three selected frequencies. Clearly, $(f_3 + f_2)/2$ (1, black), $f_5$ (2, red) and $f_7$ (3, blue) exhibit power law dependences with slopes $m_1 = 0.85 \pm 0.1$, $m_2 = 0.7 \pm 0.1$ and $m_3 = 0.75 \pm 0.1$, respectively. These exponents are greater than $\Delta E_{max} \propto P_{rf}^{1/2}$ for linear deformation potential coupling, as observed in similar hybrid devices for propagating SAWs[29,30]. It is also less than $\Delta E_{max} \propto P_{rf}^1$, expected for Stark-effect modulation, which however, can be excluded in our device.

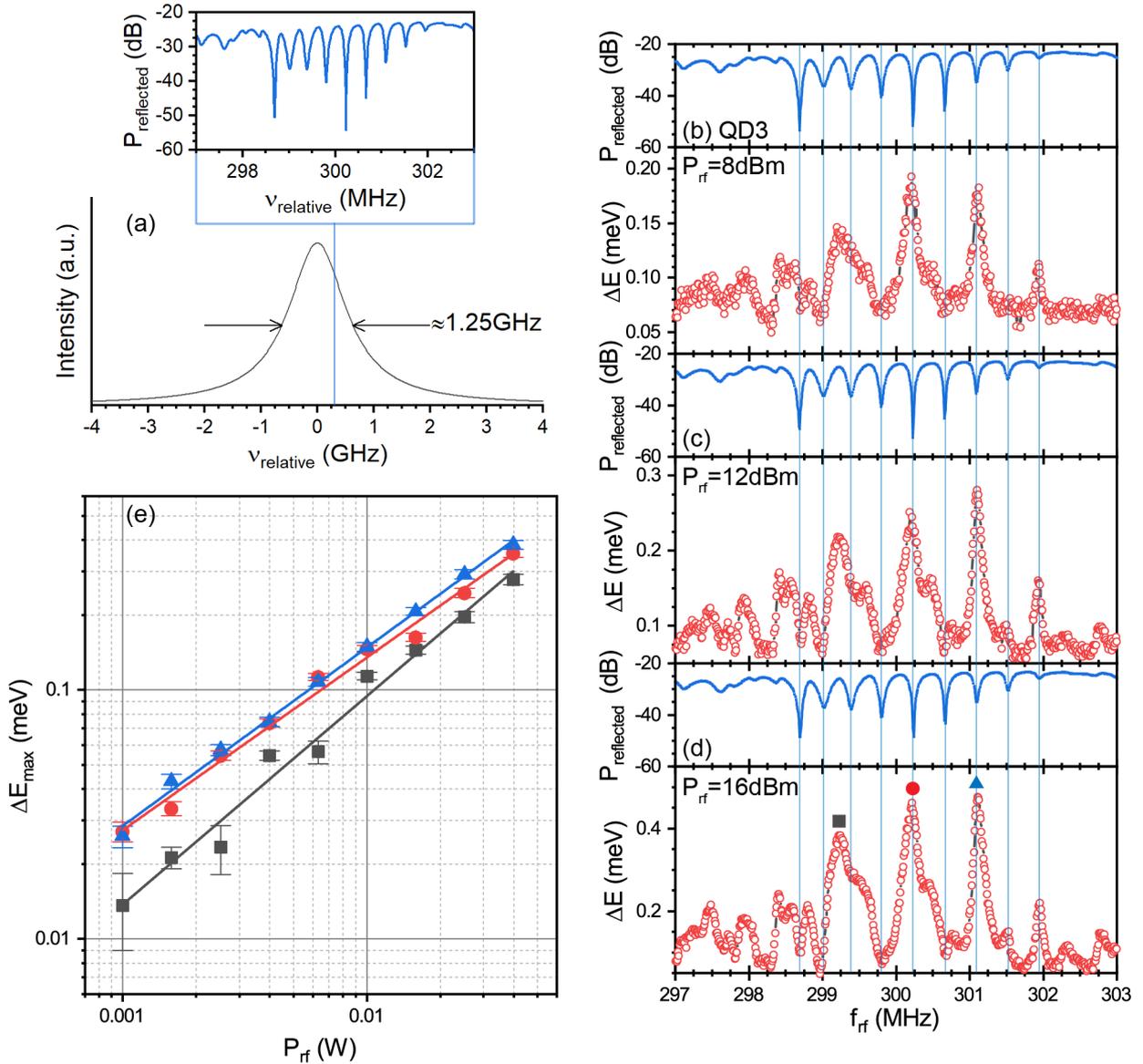

*Figure 4 – (a) Schematic showing the line width of the QD in comparison to the phononic spectrum of the resonator. (ab-d) Optomechanical response of QD3 for three selected values of $P_{rf}$ 8 (b), 12 (c) and 16 dBm (d) (main panels). Upper panels show the simultaneously measured $P_{reflected}$. (e) Amplitudes of three selected peaks [marked by corresponding symbols in (d)] as a function of $P_{rf}$.*

In conclusion, we demonstrate the heterointegration of an (Al)GaAs based QD-heterostructure on a LiNbO$_3$ SAW resonator. In our hybrid device we demonstrate strong optomechanical coupling between single QDs with the phononic modes of the SAW-resonator. This coupling is position dependent determined by the spatial overlap of the acoustic field and the QD. Our platform represents an important step towards hybrid semiconductor-LiNbO$_3$ quantum devices. In particular, our approach is fully compatible with emerging thin film LiNbO$_3$ technology[38–41] and a wide variety of quantum emitters[42]. Moreover, it can be readily combined with electrical contacts[30] facilitating quasi-static Stark-tuning of the QD's optical transitions. Finally, small mode volume and high frequency ($> 1\,\mathrm{GHz}$) resonators may enable coherent optomechanical control in the resolved sideband regime which has been reached both for III-V QDs[13,43] and defect centers[10]. These advancements may allow to unravel the physical origin of the complex optomechanical coupling observed for certain frequencies and at high acoustic drive amplitudes. Finally, the demonstrated hybrid architecture promises a strong enhancement of the optomechanical coupling compared to traditional monolithic approaches[44].

**Supplementary material**
See the supplementary material for the sample design, details on the optical experiments, rf characterization, FEM, additional experimental data of QD5, and best fits of the data in Figure 4.

**Acknowledgements**
This project has received funding from the European Union's Horizon 2020 research and innovation programme under the Marie Sklodowska-Curie grant agreement No 642688 (SAWtrain), and the Deutsche Forschungsgemeinschaft (DFG) via the Cluster of Excellence "Nanosystems Initiative Munich" (NIM), the Austrian Science Fund (FWF): P 29603, I 4320, the Linz Institute of Technology (LIT) and the LIT Lab for secure and correct systems, supported by the State of Upper Austria. We thank Maximilian Gnedel and Ferdinand Haider for performing TEM and Achim Wixforth for his continuous support and invaluable discussions.

**Data availability**
The data that support the findings of this study are available from the corresponding author upon reasonable request.

# Supplementary Material for:
# "A Hybrid (Al)GaAs-LiNbO$_3$ Surface Acoustic Wave Resonator for Cavity Quantum Dot Optomechanics"

Emeline D. S. Nysten, Armando Rastelli, Hubert J. Krenner[*]
[*]hubert.krenner@physik.uni-augsburg.de

### i. Sample design

The semiconductor membrane was grown by molecular beam epitaxy on top of a 1µm-thick Al$_{0.75}$Ga$_{0.25}$As sacrificial layer. The membrane consists of 140 nm thick Al$_{0.33}$Ga$_{0.67}$As layer sandwiched between 5-nm-thick GaAs passivation layers amounting to a total thickness of 150 nm. In the center of this membrane, a layer of GaAs QDs was fabricated by a droplet etching and filling technique[1]. Epitaxial lift-off and transfer of the (Al)GaAs structure was performed by selectively etching the sacrificial layer in a dilute HF solution[2–5]. The membrane was subsequently transferred onto the Pd layer. A strong mechanical bond forms at the interface between the metal and the III-V semiconductor[6]. This is crucial for faithful transduction of the acoustic field into the semiconductor. From the as-transferred film, a rectangular-shaped part was isolated by wet chemical etching. A microscope image of the final device is shown in SFig 1. It shows the two Bragg mirrors on the left and on the right are separated by approximately 4.5 mm. The driving IDT is positioned off-center close to the right Bragg mirror. The center part of the resonator is covered by a 3mm long Pd adhesion layer with the 215 µm (Al)GaAs membrane close to its center.

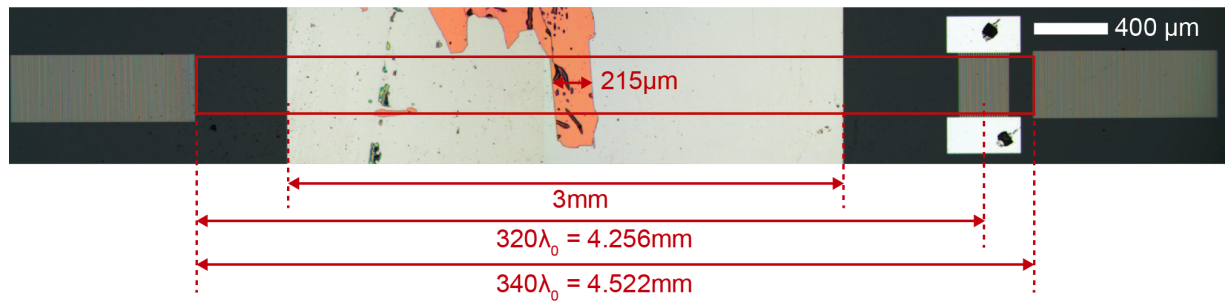

SFig 1 – Microscope image of the hybrid device.

### ii. Details on optical experiments

A schematic of the full experimental setup is shown in SFig 2. The reported optical experiments were performed in a liquid helium flow cryostat in a conventional micro-photoluminescence (µ-PL) setup (lower right part of SFig 2). A pulsed diode laser (wavelength 660 nm) emitting 90 ps long pulses with a repetition rate of 80 MHz is focused to a diffraction limited spot (diameter $\simeq 1.5\ \mu m$) on the sample to generate electrons and holes. The surface density of these QDs was $< 1\ \mu m^{-2}$, which allows to isolate individual QDs. We detect the time-integrated PL emission of single QDs as a function of the applied electrical $f_{rf}$ by a 0.5 m imaging monochromator equipped with a cooled CCD detector. In the studied frequency range $285\ \mathrm{MHz} \leq f_{rf} \leq 315\ \mathrm{MHz}$ the laser repetition rate and the electrical frequency are not commensurate, i.e. $f_{rf} \neq m \cdot 80\ \mathrm{MHz}$, $m$ integer. Thus, the observed spectral broadening is a measure for the amplitude of the optomechanical modulation[5,7].

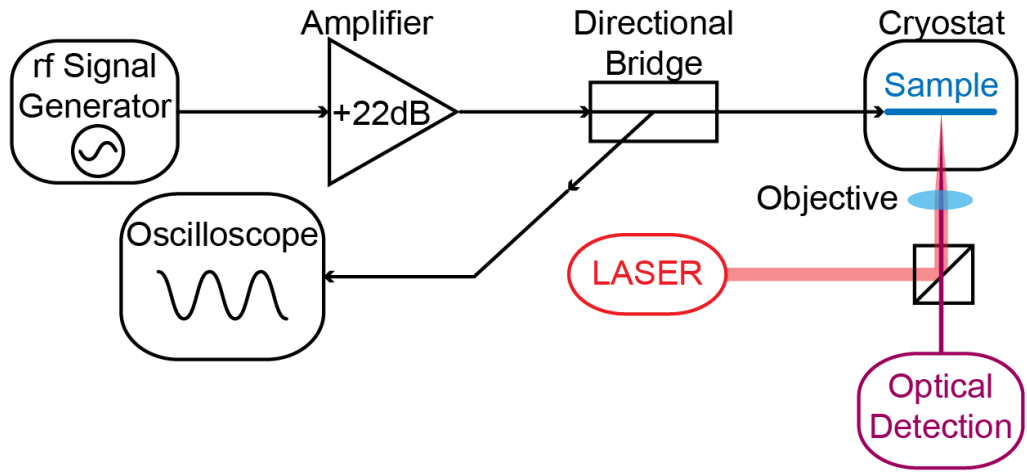

*SFig 2 – Schematic of the experimental setup.*

The electrical part of the setup shown in SFig 2 allows to apply a rf voltage to the IDT and detect the electrical power reflected. The output of the rf signal generator is amplified and connected via a directional bridge to the sample inside the cryostat. The reflected electrical signal is sent from the directional bridge to an oscilloscope. Thus, the reflected power $P_{reflected}$ is measured simultaneously for any combination of electrical frequency ($f_{rf}$) and power ($P_{rf}$).

### iii. Room temperature rf characterization

The room temperature data in Figure 1 of the main letter were measured with a standard vector network analyzer and were fitted using a model put forward by Manenti and coworkers[8]. The obtained key parameters, $f_n$, $Q_{i,n}$ and the mode splitting $\Delta f = f_{n+1} - f_n$ are summarized in STab 1 and STab 2 for the bare SAW resonator and the full hybrid device, respectively. The mean and standard deviation stated in the main letter are calculated from these data and given in the main letter.

*STab 1 – Bare SAW resonator before heterointegration*

|  | 1 | 2 | 3 | 4 | 5 | 6 | 7 | 8 | 9 |
|---|---|---|---|---|---|---|---|---|---|
| $n_{abs}$ | 708 | 709 | 710 | 711 | 712 | 713 | 714 | 715 | 716 |
| $f_n$ (MHz) | 294.40 | 294.76 | 295.14 | 295.56 | 295.99 | 296.42 | 296.85 | 297.28 | 297.71 |
| $Q_{i,n}$ | 4705 | 2412 | 2469 | 2310 | 2827 | 2781 | 2939 | 2897 | 2727 |
| $\Delta f$ (MHz) | 0.358 | 0.387 | 0.421 | 0.424 | 0.432 | 0.429 | 0.431 | 0.434 | |

*STab 2 – Hybrid device*

|  | 1 | 2 | 3 | 4 | 5 | 6 | 7 | 8 |
|---|---|---|---|---|---|---|---|---|
| $n_{abs}$ | 727 | 728 | 729 | 730 | 731 | 732 | 733 | 734 |
| $f_n$ (MHz) | 294.38 | 294.74 | 295.13 | 295.54 | 295.96 | 296.38 | 296.80 | 297.22 |
| $Q_{i,n}$ | 2302 | 1979 | 2291 | 2483 | 2666 | 2700 | 2694 | 2840 |
| $\Delta f$ (MHz) | 0.359 | 0.398 | 0.409 | 0.419 | 0.423 | 0.419 | 0.414 | |

## iv. Finite element modelling

We studied the acoustic properties of the heterointegrated (Al)GaAs-Pd-LiNbO$_3$ structure employing finite element modelling as established in Ref. [5].

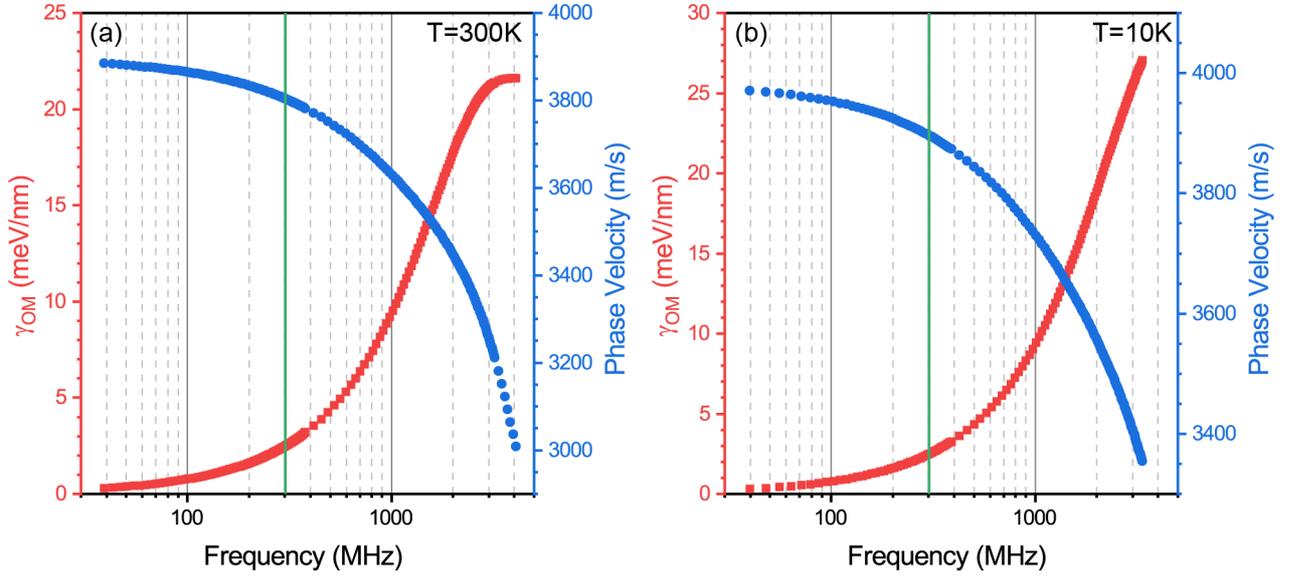

*SFig 3 – FEM simulation of the optomechanical coupling parameter $\gamma_{om}$ (red) and SAW phase velocity $c_{SAW,hybrid}$ (blue) of the heterointegrated structure at (a) T=300K and (b) T=10K.*

SFig 3 shows in (blue) the calculated phase velocity $c_{SAW,hybrid}$ of the full stack as a function of acoustic frequency $f_{SAW}$. Analogous calculations (not shown) have been performed for the LiNbO$_3$ surface coated with 50 nm of Pd without the semiconductor. In addition, SFig 3 shows in (red) the optomechanical coupling parameter $\gamma_{om}$ for QDs located in the center of the membrane.

From our FEM we obtain a phase velocity at $f_{SAW} = 300$ MHz on the bare LiNbO$_3$ surface of $c_{SAW,0} = 3990$ m/s. Pd and the (Al)GaAs heterostructure are acoustically slow materials. Thus, the phase velocity reduces to $c_{SAW,LiNiO_3+Pd} = 3830\,\frac{m}{s}$ and $c_{SAW,LiNbO_3+Pd+(Al)GaAs} = 3800\,\frac{m}{s}$ in the Pd-coated and fully heterointegrated regions, respectively.

Based on the such calculated phase velocities the cavity roundtrip time at $T$ = 300K is given by $T_c = 2 \cdot \frac{4800 \pm 50\,\mu m}{3990\,\frac{m}{s}} = 2.41s \pm 0.025\,\mu s$ for the bare resonator and $T_{c,hybrid} = 2 \cdot \left( \frac{(4800 \pm 50\,\mu m) - 3000\,\mu m - 215\,\mu m}{3990\,\frac{m}{s}} + \left(\frac{3000\,\mu m - 215\,\mu m}{3830\,\frac{m}{s}}\right)_{Pd} + \left(\frac{215\,\mu m}{3800\,\frac{m}{s}}\right)_{full\,stack} \right) = 2.47 \pm 0.025\,\mu s$ for the heterointegrated resonator. The $\Delta T_c \simeq 60$ ns is in excellent agreement with the experimental characterization.

We note that similar considerations could be made also for the contribution by the IDT. Since resonator and IDT are fabricated in a single step, an analogous analysis is not possible.

### v. Additional data from a quantum dot in a second device

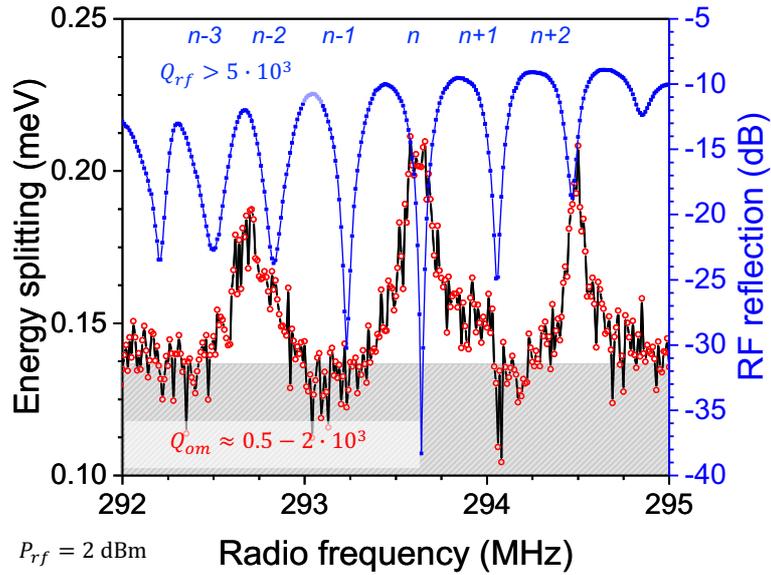

*SFig 4 – Optomechanical response of a single QD, QD5 (red symbols/black line) and simultaneously measured electrical reflection (blue) in a different hybrid device.*

SFig 4 shows additional data from a second device. The resonator is of nominally identical design as that presented in the main letter and shown above. Both the electrically measured mode spectrum in the reflected rf power (blue) and Q-factors and the optomechanical response of a single QD, QD5 (red symbols) are nearly identical to those presented in the main letter. Because the width of the QD heterostructure was slightly different, the center of the mode is shifted to lower frequencies, but the overall pattern and high Q-factors are nicely preserved. In these data, QD5 couples strongly to the center mode and at the low frequency side of the mode pattern, a clear optomechanical response is again detected exactly in the center between to electrically detected $n-3$ and $n-2$ modes. The observed behavior faithfully reproduces that observed for QD3 in the main letter.

## vi. Fitting procedure of data in Figure 4

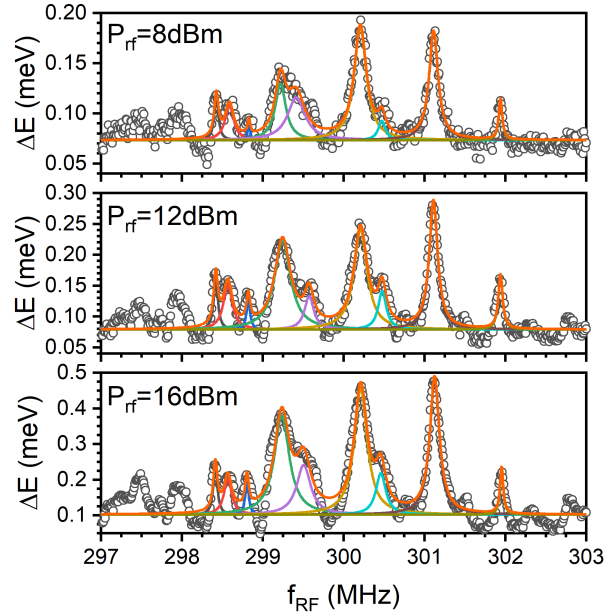

*SFig 5 – Best fits (lines) to the data (symbols) in Figure 4.*

The optomechanical response of the QD $\Delta E(f_{rf})$ presented in Figure 4 of the main letter is fitted as a series of Lorentzian lines. SFig 5 shows the experimental data (symbols) and the best fit (lines). The area extracted for selected peaks is plotted in Figure 4 (e).